# Nothing but Relativity, Redux[†]

*Joel W. Gannett*
Senior Scientist

Telcordia Technologies, Inc., 331 Newman Springs Road, Red Bank, NJ 07701 USA
Email: jgannett@research.telcordia.com

## Abstract

A previous paper demonstrated that spacetime transformations consistent with the principle of relativity can be derived without assuming explicitly the constancy of the speed of light. Here we correct an error in the earlier paper while showing that this derivation can be done under weaker assumptions, in particular, without an implicit assumption of differentiability or even continuity for the spacetime mapping. Hence, these historic results could have been derived centuries ago, even before the advent of calculus.

Relativity—the notion that the laws of physics remain the same in all inertial frames—is a cornerstone of Einstein's Special Theory of Relativity [1] as well as a cornerstone of physics in general. The other cornerstone of Einstein's Relativity is the notion that the speed of light remains constant in all inertial frames. Einstein postulated the constancy of the speed of light and showed how that brought electrodynamic phenomena under the purview of the principle of relativity.

In his thought-provoking paper titled *Nothing but Relativity*, Palash B. Pal showed how spacetime transformations consistent with the Einsteinian (Lorentz) transformation can be derived without assuming explicitly the constancy of the speed of light and without invoking gedanken experiments involving light rays [2]. This observation, which may seem remarkable to many, has been noted by numerous authors dating as far back as 1910. Papers by Ignatowsky [3, 4] and Frank and Rothe [5] are notable early examples.[1] Pal [2] lists authors who published variations on this theme going back to the mid-1960s, while Berzi and Gorini's 1969 paper [6] lists many earlier references. Some physics textbooks reflect this "relativity only" observation [7], while others [8] follow Einstein's seminal paper [1] by basing the derivation on an assumed constancy of the speed of light.

The straightforward mathematical derivation given by Pal [2] was based on the principle of relativity, the isotropy of space, and the homogeneity of space and time. Pal invoked the principle of relativity only insofar as needed to assert the reciprocity property of the inertial spacetime transformation, which Berzi and Gorini [6] showed was a nontrivial consequence of the principle of relativity and the isotropy of space. We point out here that a less restrictive but adequate set of assumptions for deriving these results can be obtained if one bypasses the principle of relativity altogether and instead assumes reciprocity directly, in which case the current paper might have been titled *Nothing but Relativity, and Not That Either*, or perhaps *Nothing but Reciprocity*. At any rate, working from his assumptions, Pal [2] showed that one can deduce two classes of transformations: Classical Galilean, in which time is the same in all inertial frames, and the Einsteinian "universal speed limit" solution, in which time differs between two inertial frames in nonzero relative motion. The final selection between the two must favor the Einsteinian, since (as remarked above) this spacetime view brings electrodynamic phenomena under the purview of the principle of relativity when the universal speed limit is identified as the speed of light.

---

[†] *Eur. J. Phys.* 28 (2007) 1145-1150.
[1] Ignatowsky's paper in vol. 17 of *Archiv Der Mathematik Und Physik* [4] has been erroneously referenced as published in 1910, while his vol. 18 paper in the same journal has been referenced as published in 1911. In fact, both vols. 17 and 18 were published in 1911 (vol. 16 covered 1910 and vol. 19 covered 1912).

Pal's derivation [2] incorporated an implicit assumption that the transformation mapping was differentiable, since a step in his argument proving linearity from homogeneity involved taking the derivative of this mapping. Einstein may have had the same assumption in mind when he wrote his seminal 1905 paper [1], although it is not evident what Einstein's assumptions were in this regard since he merely stated that homogeneity clearly implied linearity but did not show any steps in his reasoning. Here we show that one can drop the differentiability assumption and proceed by assuming merely that the transformation is bounded on a compact set. Hence, Einsteinian relativity could have been derived centuries ago, even before the advent of calculus. In this paper we also sidestep an apparent error in Pal's original derivation [2].

Let $S$ denote our reference frame with spacetime coordinates denoted $(x, y, z, t)$. Let $S'$ denote a frame moving parallel to the $x$-axis with constant speed $v$ relative to $S$. Denote the spacetime coordinates of $S'$ by $(x', y', z', t')$. We will follow Pal [2] by assuming that the spatial origins of the two coordinate systems coincide at $t = 0$; that is, $x = y = z = t = 0$ implies $x' = y' = z' = t' = 0$.

We seek a transformation, with properties to be discussed, that maps the spacetime coordinates of $S$ to the spacetime coordinates of $S'$ and has the following form:[2]

$$x' = X(x, y, z, t, v) \qquad (1)$$

$$y' = Y(x, y, z, t, v) \qquad (2)$$

$$z' = Z(x, y, z, t, v) \qquad (3)$$

$$t' = T(x, y, z, t, v) \qquad (4)$$

where $X(\cdot)$, $Y(\cdot)$, $Z(\cdot)$, and $T(\cdot)$ are the coordinate functions mapping $\mathbf{R}\times\mathbf{R}\times\mathbf{R}\times\mathbf{R}\times\mathbf{R}$ to $\mathbf{R}$. Here $\mathbf{R}$ denotes the set of real numbers.

We start by proving that the desired mapping indicated in Eqs. (1-4) is linear. Here we mention our point of disagreement with the argument given by Pal [2], but the problem is easily overcome, as we shall demonstrate. Pal [2] considered a rigid rod of length $l$ that is stationary in $S$ and placed along the $x$ axis between points $x_1$ and $x_2$, where of course $l = |x_2 - x_1|$. At a particular time $t_0$, he considered the difference in the $x'$ coordinates of the endpoints, which he denoted $l'$, as follows:

$$l' = X(x_2, 0, 0, t_0, v) - X(x_1, 0, 0, t_0, v) \qquad (5)$$

---

[2] Pal [2] apparently assumed tacitly, without derivation, that $y'=y$ and $z'=z$, and then he dropped these variables from his equations without further consideration or comment. Einstein [1] provided a proof that $y'=y$ and $z'=z$, but he invoked a gedanken experiment involving light rays, and he used calculus as well. We choose here to include $y$ and $z$ in our non-gedanken, non-calculus analysis.

Pal [2] asserted that this difference represents the length of the rod in $S'$, but we disagree because the two endpoints of the rod will in general map at $t_0$ into two different $t'$ times, to wit: $t'_1 = T(x_1, 0, 0, t_0, v)$ and $t'_2 = T(x_2, 0, 0, t_0, v)$. We would need to calculate the end points at the *same t'* time to find the length of the rod in $S'$, since a rigid rod's length in a frame of reference is defined as the distance between its end points at the same time instant, where time is reckoned according to that frame. Indeed, if one uses the Lorentz transformation in Eq. (5), one obtains the erroneous result $l' = \gamma(x_2 - x_1)$, where $\gamma = 1/\sqrt{1 - v^2/c^2}$ is the Lorentz factor. The correct answer is $l' = (x_2 - x_1)/\gamma$, which shows the well-known length contraction phenomenon.

Although we cannot tie the coordinate difference in Eq. (5) directly to the length of a physical object, the basic idea behind Eq. (5)—that of investigating how intervals map between the reference frames—points the way to proving linearity.[3] Let $x_0, y_0, z_0, t_0, \Delta x, \Delta y, \Delta z,$ and $\Delta t$ denote arbitrary real numbers, and define $\Delta x', \Delta y', \Delta z'$, and $\Delta t'$ as follows:

$$\Delta x' = X(x_0 + \Delta x, y_0 + \Delta y, z_0 + \Delta z, t_0 + \Delta t, v) - X(x_0, y_0, z_0, t_0, v) \qquad (6)$$

$$\Delta y' = Y(x_0 + \Delta x, y_0 + \Delta y, z_0 + \Delta z, t_0 + \Delta t, v) - Y(x_0, y_0, z_0, t_0, v) \qquad (7)$$

$$\Delta z' = Z(x_0 + \Delta x, y_0 + \Delta y, z_0 + \Delta z, t_0 + \Delta t, v) - Z(x_0, y_0, z_0, t_0, v) \qquad (8)$$

$$\Delta t' = T(x_0 + \Delta x, y_0 + \Delta y, z_0 + \Delta z, t_0 + \Delta t, v) - T(x_0, y_0, z_0, t_0, v) \qquad (9)$$

Clearly, the 4-tuple $(\Delta x, \Delta y, \Delta z, \Delta t)$ represents the coordinates in a system $\hat{S}$ that is stationary with respect to $S$ and identical to $S$ except that the origin of $\hat{S}$ appears at $(x_0, y_0, z_0, t_0)$ in the $S$ system. Likewise, the 4-tuple $(\Delta x', \Delta y', \Delta z', \Delta t')$ represents the coordinates in a system $\hat{S}'$ that is stationary with respect to $S'$ and identical to $S'$ except that the origin of $\hat{S}'$ appears at $(x'_0, y'_0, z'_0, t'_0)$ in the $S'$ system, where $x'_0 = X(x_0, y_0, z_0, t_0, v)$, $y'_0 = Y(x_0, y_0, z_0, t_0, v)$, etc. Note that $\hat{S}$ and $\hat{S}'$ bear the same relationship to each other as the systems $S$ and $S'$ bear to each other; in particular, $(\Delta x, \Delta y, \Delta z, \Delta t) = (0, 0, 0, 0)$ implies $(\Delta x', \Delta y', \Delta z', \Delta t') = (0, 0, 0, 0)$. Consequently, by the homogeneity of space and time, $\hat{S}$ and $\hat{S}'$ are related by exactly the same mapping functions shown in Eqs. (1-4):

---

[3] While revising this paper in response to referee comments, the author discovered that this proof of linearity is similar to a proof by Berzi and Gorini [6]. However, Berzi and Gorini assumed continuity at the origin, while our proof proceeds from a weaker boundedness assumption and no assumption of continuity.

$$\Delta x' = X(\Delta x, \Delta y, \Delta z, \Delta t, v) \tag{10}$$

$$\Delta y' = Y(\Delta x, \Delta y, \Delta z, \Delta t, v) \tag{11}$$

$$\Delta z' = Z(\Delta x, \Delta y, \Delta z, \Delta t, v) \tag{12}$$

$$\Delta t' = T(\Delta x, \Delta y, \Delta z, \Delta t, v) \tag{13}$$

Combining Eqs. (6) and (10), we obtain

$$X(x_0 + \Delta x, y_0 + \Delta y, z_0 + \Delta z, t_0 + \Delta t, v) = X(x_0, y_0, z_0, t_0, v) + X(\Delta x, \Delta y, \Delta z, \Delta t, v) \tag{14}$$

with similar equations for $Y$, $Z$, and $T$ also holding.

For notational convenience, define a function $\mathbf{f}_v(\cdot)$ mapping $\mathbf{R}^4$ to $\mathbf{R}^4$ whose component functions are the four functions in Eqs. (1-4) with $v$ set to the indicated subscript value and with the domain of $\mathbf{f}_v(\cdot)$ being the set of 4-vectors $\mathbf{w} \in \mathbf{R}^4$ whose components are $x$, $y$, $z$, and $t$. Then Eqs. (1-4) can be rewritten

$$\mathbf{w}' = \mathbf{f}_v(\mathbf{w}) \tag{15}$$

The assumption that $\mathbf{f}_v(\cdot)$ maps between two systems whose origins coincide in 4-space means that $\mathbf{f}_v(\mathbf{0}) = \mathbf{0}$. If $\mathbf{w}$ and $\mathbf{u}$ are arbitrary elements of $\mathbf{R}^4$, then Eq. (14) along with its corresponding equations for $Y$, $Z$, and $T$ gives

$$\mathbf{f}_v(\mathbf{w} + \mathbf{u}) = \mathbf{f}_v(\mathbf{w}) + \mathbf{f}_v(\mathbf{u}) \tag{16}$$

Since $\mathbf{f}_v(\mathbf{0}) = \mathbf{0}$, it follows from Eq. (16) with $\mathbf{u} = -\mathbf{w}$ that

$$\mathbf{f}_v(-\mathbf{w}) = -\mathbf{f}_v(\mathbf{w}) \tag{17}$$

From these properties, it is easy to show that for any integer $m$,

$$\mathbf{f}_v(m\mathbf{w}) = m\mathbf{f}_v(\mathbf{w}) \tag{18}$$

To prove Eq. (18), note that it is true trivially for $m = 0$ or $m = 1$. Suppose Eq. (18) is true for some nonnegative integer $m$. We prove Eq. (18) by induction using Eq. (16): $\mathbf{f}_v((m+1)\mathbf{w}) = \mathbf{f}_v(m\mathbf{w} + \mathbf{w}) = \mathbf{f}_v(m\mathbf{w}) + \mathbf{f}_v(\mathbf{w}) = m\mathbf{f}_v(\mathbf{w}) + \mathbf{f}_v(\mathbf{w}) = (m+1)\mathbf{f}_v(\mathbf{w})$. If $m < 0$, let $p = -m$. Note that $\mathbf{f}_v(p\mathbf{w}) = p\mathbf{f}_v(\mathbf{w})$, as we have just proved. Use Eq. (17) to get $\mathbf{f}_v(-p\mathbf{w}) = -p\mathbf{f}_v(\mathbf{w})$. Finally, obtain Eq. (18) for this case by substituting $m = -p$.

For any nonzero integer $n$ and any $\mathbf{u} \in \mathbf{R}^4$, substitute $m = n$ and $\mathbf{w} = \mathbf{u}/n$ in Eq. (18) to obtain

$$\mathbf{f}_v\left(\frac{\mathbf{u}}{n}\right) = \frac{1}{n}\mathbf{f}_v(\mathbf{u}) \tag{19}$$

If $q$ is any rational number, then it can be written in the form $m/n$ for some integers $m$ and $n$, where $n$ is nonzero. Combining Eqs. (18) and (19) gives, for any $\mathbf{w} \in \mathbf{R}^4$,

$$\mathbf{f}_v(q\mathbf{w}) = \mathbf{f}_v\left(\frac{m}{n}\mathbf{w}\right) = m\mathbf{f}_v\left(\frac{1}{n}\mathbf{w}\right) = \frac{m}{n}\mathbf{f}_v(\mathbf{w}) = q\mathbf{f}_v(\mathbf{w}) \qquad (20)$$

For any integer $n > 0$, let $B_{\frac{1}{n}}$ denote the closed Euclidean ball in $\mathbf{R}^4$ of diameter $1/n$. That is, $B_{\frac{1}{n}} = \{\mathbf{w} \in \mathbf{R}^4 : \|\mathbf{w}\| \leq \frac{1}{n}\}$, where $\|\circ\|$ denotes the Euclidean norm. Our boundedness assumption is that $\mathbf{f}_v(\cdot)$ is bounded on $B_1$. That is, we assume there exists a real number $M_v$ such that $\|\mathbf{f}_v(\mathbf{w})\| \leq M_v$ for all $\mathbf{w} \in B_1$. From this assumption we prove that $\mathbf{f}_v(\cdot)$ is continuous everywhere. Indeed, if $\mathbf{w}_n \in B_{\frac{1}{n}}$ for some integer $n > 0$, then $n\mathbf{w}_n \in B_1$. Consequently, $\|\mathbf{f}_v(n\mathbf{w}_n)\| \leq M_v$, and therefore from Eq. (20) and the properties of the Euclidean norm it follows that $\|\mathbf{f}_v(\mathbf{w}_n)\| \leq \frac{M_v}{n}$. Because the right-hand side of this inequality goes to zero as $n \to \infty$, and since $\mathbf{f}_v(\mathbf{0}) = \mathbf{0}$, it follows that $\mathbf{f}_v(\cdot)$ is continuous at $\mathbf{w} = \mathbf{0}$. From Eq. (16), continuity at $\mathbf{0}$ implies that $\mathbf{f}_v(\cdot)$ is continuous everywhere. This follows by setting $\mathbf{u} = \Delta\mathbf{w}$ in Eq. (16) and letting $\Delta\mathbf{w} \to \mathbf{0}$.

Finally, let $s$ denote any real number, possibly irrational. Let $q_i$ denote a sequence of rational numbers that approaches $s$ arbitrarily closely, that is, $\lim_{i \to \infty} q_i = s$. From the continuity of $\mathbf{f}_v(\cdot)$, which we have just proved, it follows that for any $\mathbf{w} \in \mathbf{R}^4$

$$\mathbf{f}_v(s\mathbf{w}) = \mathbf{f}_v(\lim_{i \to \infty} q_i \mathbf{w}) = \lim_{i \to \infty} \mathbf{f}_v(q_i \mathbf{w}) = \lim_{i \to \infty} q_i \mathbf{f}_v(\mathbf{w}) = s\mathbf{f}_v(\mathbf{w}) \qquad (21)$$

Eq. (21) implies that $\mathbf{f}_v(\cdot)$ is a linear homogeneous function. Indeed, let $w_i$, $1 \leq i \leq 4$, denote the four real-valued components of $\mathbf{w} \in \mathbf{R}^4$, and let $\mathbf{e}_i$, $1 \leq i \leq 4$, denote the four basis vectors in $\mathbf{R}^4$. Then by Eqs. (16) and (21),

$$\mathbf{f}_v(\mathbf{w}) = \mathbf{f}_v(w_1\mathbf{e}_1 + w_2\mathbf{e}_2 + w_3\mathbf{e}_3 + w_4\mathbf{e}_4) = \mathbf{f}_v(\mathbf{e}_1)w_1 + \mathbf{f}_v(\mathbf{e}_2)w_2 + \mathbf{f}_v(\mathbf{e}_3)w_3 + \mathbf{f}_v(\mathbf{e}_4)w_4 \qquad (22)$$

Hence, Eqs. (1-4) can be represented in matrix notation as

$$\begin{bmatrix} x' \\ y' \\ z' \\ t' \end{bmatrix} = \begin{bmatrix} h_{11} & h_{12} & h_{13} & h_{14} \\ h_{21} & h_{22} & h_{23} & h_{24} \\ h_{31} & h_{32} & h_{33} & h_{34} \\ h_{41} & h_{42} & h_{43} & h_{44} \end{bmatrix} \begin{bmatrix} x \\ y \\ z \\ t \end{bmatrix} \qquad (23)$$

where each of the matrix elements $h_{ij} = f_{v,i}(\mathbf{e}_j)$ is possibly a function of $v$, and $f_{v,i}(\cdot)$ represents the $i^{\text{th}}$ component function of $\mathbf{f}_v(\cdot)$.

Consider how the $y$ and $z$ coordinates affect $x'$ and $t'$. Suppose the $S$ and $S'$ coordinate systems are rotated 180 degrees about the $x$ axis so that the same physical events have their $y$ and $z$ coordinates reversed in sign in the rotated versus the non-rotated systems.

By the isotropy of space, the same matrix shown in Eq. (23) specifies the transformation between the rotated reference frames. But this rotation cannot affect the *x'* and *t'* coordinates, as the physical events are the same and the *x'* and *t'* axes are not affected by this rotation. This requirement can be met by Eq. (23) if and only if $h_{12} = h_{13} = h_{42} = h_{43} = 0$.

By rotating the *S* and *S'* coordinate systems 180 degrees about their *y* axes, we make the *x* and *z* axes point in the opposite direction relative to their corresponding unrotated systems. Since the *y* coordinates of events could not be affected by such a rotation, we conclude that $h_{21} = h_{23} = 0$.

Likewise, by rotating the *S* and *S'* coordinate systems 180 degrees about their *z* axes, we make the *x* and *y* axes point in the opposite direction relative to their corresponding unrotated systems. Since the *z* coordinates of events could not be affected by such a rotation, we conclude that $h_{31} = h_{32} = 0$.

Consider $h_{24}$ and $h_{34}$. An event occurring at position $(x,y,z) = (0,0,0)$ at some time *t* in the *S* frame would map to *y'* coordinate $y' = h_{24}t$ and *z'* coordinate $z' = h_{34}t$ in the *S'* frame. By the isotropy of space, these same *h* coefficients must apply if both coordinate systems are rotated 180 degrees about their *x* axes, yet this rotation would also imply that $y' = -h_{24}t$ and $z' = -h_{34}t$. These two requirements are satisfied if and only if $h_{24} = h_{34} = 0$.

For the matrix elements affecting or affected by *y*, *z*, *y'*, and *z'*, this leaves only $h_{22}(v)$ and $h_{33}(v)$ as nonzero, where here we are showing the possible dependence on *v* explicitly. Following Einstein [1], consider a stationary rod of length *L* placed between $(x,y,z) = (0,0,0)$ and $(0,L,0)$ in the *S* system. The length of the rod in the *S'* system is then $|Lh_{22}(v)|$. This length must be independent of the direction of *v*, whence we conclude that $h_{22}(v) = h_{22}(-v)$. On the other hand, applying Eq. (23) twice to convert first to *S'* and then to a new system *S''* that is stationary with respect to *S*, we conclude that $h_{22}(v)h_{22}(-v) = 1$. From these two properties, it follows that either $h_{22}(v) = 1$, independent of *v*, or $h_{22}(v) = -1$, independent of *v*.[4] We assume the concept of "direction" is the same in both the *S* and *S'* coordinate systems, and that the *y* axes of both systems point in the same direction. Since the rod of length *L*, when traversed from its end at the origin to its other end, defines the direction of the *y* axis in *S*, we can discard the −1 solution since the direction of this traversal must be the same in *S'*. Consequently, $h_{22}(v) = 1$, independent of *v*. Likewise, we conclude that $h_{33}(v) = 1$, independent of *v*.

Converting to Pal's notation [2] for the corner matrix elements, we have shown that Eq. (23) reduces to

$$\begin{bmatrix} x' \\ y' \\ z' \\ t' \end{bmatrix} = \begin{bmatrix} A_v & 0 & 0 & B_v \\ 0 & 1 & 0 & 0 \\ 0 & 0 & 1 & 0 \\ C_v & 0 & 0 & D_v \end{bmatrix} \begin{bmatrix} x \\ y \\ z \\ t \end{bmatrix} \tag{24}$$

where possible $v$ dependence is indicated by $v$ subscripts on the corner elements.

---

[4] Einstein [1] ignored the −1 solution without explanation.

The four properties of the mapping functions listed below result as immediate consequences of reciprocity and the isotropy of space. Here we have dropped the $y$ and $z$ dependence to conform to Pal's notation [2].

$$x = X(X(x,t,v), T(x,t,v), -v) \tag{25}$$

$$t = T(X(x,t,v), T(x,t,v), -v) \tag{26}$$

$$X(-x,t,-v) = -X(x,t,v) \tag{27}$$

$$T(-x,t,-v) = T(x,t,v) \tag{28}$$

Using Eqs. (24-28), Pal [2] showed how to derive the Einsteinian (Lorentz) transformation and the velocity addition law using, from this point forward, only algebra and basic physical considerations.

We have shown a path for deriving the Lorentz transformation and velocity addition law without an explicit assumption of the constancy of the speed of light, without gedanken experiments involving light rays, and without calculus. Two key topological concepts we did use were the density of the set of rational numbers within the set of real numbers, and a boundedness assumption from which we derived the continuity of the spacetime mapping. These density and continuity concepts were embodied in Eq. (21). The former concept means, essentially, that any irrational number can be approximated to arbitrary closeness by a rational number (a number with finite or repeating decimal representation). The continuity concept means, essentially, that a function maps "close-by" points to "close-by" points. As these concepts were likely accessible to the 17th century scientific mind, we see that Galileo probably had sufficient tools at his disposal to derive Einsteinian relativity—an observation that has been made before [9]. If Galileo had derived it, however, Occam's razor would have impelled him to discard Einsteinian relativity as a needlessly complex mathematical curiosity that was not required to resolve any outstanding issues known to 17th century science.

## Acknowledgements

The anonymous review comments from the *European Journal of Physics* greatly improved this paper. Also, the author would like to acknowledge research assistance and helpful comments from Telcordia colleague Matt Goodman, who is now with the U.S. Defense Advanced Research Projects Agency. Finally, the author would like to thank Palash B. Pal for writing his excellent 2003 paper [2] that sparked the author's interest in this topic.